\documentclass[conference]{IEEEtran}
\IEEEoverridecommandlockouts
\usepackage{cite}
\usepackage{amsmath,amssymb,amsfonts}
\usepackage{algorithmic}
\usepackage{graphicx}
\usepackage{caption}
\usepackage{textcomp}
\usepackage{xcolor}
\usepackage{url}

\def\BibTeX{{\rm B\kern-.05em{\sc i\kern-.025em b}\kern-.08em
    T\kern-.1667em\lower.7ex\hbox{E}\kern-.125emX}}
\begin{document}

\title{Detecting Prompt Injection Attacks Against Application Using Classifiers\\}

\author{\IEEEauthorblockN{Safwan Shaheer}
\IEEEauthorblockA{\textit{Department of Computer Science (CS)} \\
\textit{School of Data and Sciences (SDS)}\\
Dhaka, Bangladesh \\
safwan.shaheer@g.bracu.ac.bd}
\and
\IEEEauthorblockN{G. M. Refatul Islam}
\IEEEauthorblockA{\textit{Department of Computer Science and Engineering (CSE)} \\
\textit{School of Data and Sciences (SDS)}\\
Dhaka, Bangladesh \\
gm.refatul.islam@g.bracu.ac.bd}
\and
\IEEEauthorblockN{Mohammad Rafid Hamid}
\IEEEauthorblockA{\textit{Department of Computer Science and Engineering (CSE)} \\
\textit{School of Data and Sciences (SDS)}\\
Dhaka, Bangladesh \\
mohammad.rafid.hamid@g.bracu.ac.bd}
\and
\IEEEauthorblockN{Md. Abrar Faiaz Khan}
\IEEEauthorblockA{\textit{Department of Computer Science and Engineering (CSE)} \\
\textit{School of Data and Sciences (SDS)}\\
Dhaka, Bangladesh \\
md.abrar.faiaz.khan@g.bracu.ac.bd}
\and
\IEEEauthorblockN{Md. Omar Faruk}
\IEEEauthorblockA{\textit{Department of Computer Science and Engineering (CSE)} \\
\textit{School of Data and Sciences (SDS)}\\
Dhaka, Bangladesh \\
md.omar.faruk@g.bracu.ac.bd}
\and
\IEEEauthorblockN{Yaseen Nur}
\IEEEauthorblockA{\textit{Department of Computer Science (CS)} \\
\textit{School of Data and Sciences (SDS)}\\
Dhaka, Bangladesh \\
yaseen.nur.taz@g.bracu.ac.bd}
}

\maketitle

\begin{abstract}
Prompt injection attacks are a growing concern
in various domains, as they can compromise the security and
stability of critical systems, from power grids to large-scale
web applications. In this research, we propose a comprehensive
approach for detecting and mitigating prompt injection attacks against—web applications using predefined system prompts— by highly curating a prompt injection dataset and training it
on classifiers like Long Short-Term Memory (LSTM) networks
to FNN and machine learning models such as Random Forest Classifier and Naive Bayes. We
build upon a pre-existing dataset from HuggingFace, named
HackAPrompt-Playground-Submissions. Our proposed solution aims to improve the detection and mitigation of prompt
injection attacks, ensuring the security and stability of the
targeted applications and systems.
\\
\end{abstract}


\section{Introduction}
Prompt injection attacks have emerged as a pressing and formidable threat, mounting concern due to their potential to jeopardize the security and stability of critical systems across diverse domains, encompassing essential infrastructure such as power grids and mission-critical web applications. These attacks exploit the vulnerabilities of Large Language Models (LLMs) like GPT-3.5, BERT etc. which are advanced AI models designed to generate human-like text by processing and understanding vast amounts of training data.\\
\\
Prompt injection involves maliciously manipulating or injecting harmful content into the initial prompt provided to LLMs, with the intention of influencing their generated output. This manipulation can lead to biased, misleading, or even harmful responses, compromising the integrity and reliability of decision-making processes. The impact of prompt injection attacks is far-reaching, undermining trust in AI-generated content and potentially leading to misinformation, compromised security, and harmful outcomes.\\
\\
Prompt engineering works due to the inherent nature of LLMs. These models leverage deep learning techniques to analyze patterns, context, and semantics in text, enabling them to generate coherent and contextually relevant responses. Attackers exploit this capability by carefully crafting prompts to introduce biased or malicious content, influencing the model's output in undesirable ways. \\
\\
In this research, we present an approach to effectively detect and mitigate prompt injection attacks on LLMs. To tackle the challenges posed by these attacks, we focus on curating a prompt injection dataset with various techniques. We incorporate a pre-existing dataset from HuggingFace, named HackAPrompt-Playground-Submissions. We then enhance and curate these datasets, ensuring they capture the diverse range of prompt injection attack scenarios accurately. This approach empowers our research to offer improved detection and mitigation strategy.\\
\\
Neglecting to tackle prompt injection attacks can have severe consequences. It not only compromises the reliability of AI-generated content but also perpetuates biases, hindering progress towards fairness and equity in LLMs. Our research aims to contribute to the field by providing effective detection and mitigation strategies, ultimately ensuring the responsible use and deployment of LLMs in critical systems.
\\

\section{Related Work}

In this part, we will examine some of the most important articles published on these elements and their implications for LLM-integrated applications.

\subsection{Jailbreaks in LLMs}
The research paper titled ``Tricking LLMs into Disobedience: Understanding, Analyzing, and Preventing Jailbreaks"\cite{tricking_llms} provides a formalism and taxonomy for jailbreaks in LLMs. It also surveys existing jailbreak methods and evaluates their effectiveness on open-source and commercial LLMs, such as GPT 3.5, OPT, BLOOM, and FLAN-T5-XXL. This study aims to close the existing knowledge gap regarding the understanding and analysis of these assaults and their countermeasures. To do so, the study will propose a restricted collection of quick guards and explain their success against known attack types.

\subsection{Privacy Attacks on ChatGPT}
Invasions of ChatGPT Users' PrivacyThe article ``Multi-step Jailbreaking Privacy Attacks on ChatGPT" \cite{multistep_jailbreaking} explores the privacy risks posed by OpenAI's ChatGPT and the New Bing version that ChatGPT has upgraded. The authors follow prior assaults on ChatGPT that extracted training data and suggest a multi-step jailbreaking prompt to obtain personal information. These attacks followed previous ones that extracted training data. According to the findings of the study, application-integrated LLMs have the potential to function as efficient sources of erroneous information, which can result in a considerable drop in the performance of Open-Domain Question Answering (ODQA) systems—indirect Prompt Injection.

\subsection{Indirect Prompt Injection}
 In the article ``Not what you've signed up for Compromising Real-World LLM-Integrated Applications with Indirect Prompt Injection"\cite{indirect_prompt_injection} the authors present the notion of Indirect Prompt Injection (IPI), which may be used to compromise LLM-integrated applications. The research deconstructs the intricacies and repercussions of prompt injection attacks on genuine LLM-integrated systems and creates HouYi, a unique black-box prompt injection attack approach that takes inspiration from other existing methods.

\subsection{Prompt Injection Attacks}
Attacks That Are Initiated ImmediatelyThe article ``Prompt Injection attack against LLM-integrated Applications" \cite{prompt_injection_attack} gives an in-depth study of the possible security issues that are connected with prompt injection attacks on actual LLM-integrated applications and the potential impact that these attacks may have on LLM-integrated systems. Attacks that involve prompt injection are designed to trick LLM-based software into giving replies that were not expected, to gain unauthorized access, to change results, or to fool users.

\subsection{The Significance of AI Generators in Cyber Security and Data Privacy}

The research article titled ``From ChatGPT to ThreatGPT: The Significance of AI Generative Models in Cyber Security and Data Privacy" examines the weaknesses of ChatGPT and how malevolent individuals can manipulate them to surreptitiously pilfer sensitive information while evading security measures. \cite{chatgpt_threatgpt}

\subsection{LLM-Based Web Application Defense Mechanisms}

The scholarly publication ``From Prompt Injections to SQL Injection Attacks: Evaluating the Safety of LLM-Integrated Web Applications" offers a comprehensive appraisal of prompt-to-SQL (P\_2SQL) injections targeting web-based applications that employ Langchain, revealing that unfiltered user prompts may expose said apps to SQL injection attacks, potentially compromising database security \cite{llm_webapp_protection}.

\subsection{Virtual Prompt Manipulation}
The concept of Virtual Prompt Manipulation (VPM) for guided Large Language Models (LLMs) is presented in the academic journal paper titled ``Virtual Prompt Manipulation for Guided Large Language Models"\cite{virtual_prompt_injection} .The Plug-in Model (VPM) enables an adversary to virtually embed a given command into the framework of a model. This simplifies adversarial control over the activities of the model while still maintaining the legitimacy of the initial prompt.

\subsection{Goal Hijacking and Prompt Leaking}
The article ``Ignore Previous Prompt: Attack Techniques For Language Models"\cite{ignore_previous_prompt} analyzes two different kinds of attacks, aim hijacking and prompt leakage, that are carried out against LLMs such as GPT-3. PromptInject is the name of the proposed system that the authors have developed for iterative mask-based hostile hint creation. According to the findings of this investigation, even those with low abilities but hostile intentions might use the stochastic nature of GPT-3 to produce unforeseen dangers. This conclusion was reached based on the findings of the study. Using the coding implementation of the PromptInject software program, it could be able to carry out more investigation and inspection.\\

These studies shed light on the emerging issues and challenges that are associated with the security threats that are present in LLMs. It is essential for researchers and developers to address these security issues and create effective mitigation measures in order to ensure the safe and responsible utilization of LLMs as they continue to evolve and become more incorporated into a variety of applications. This is the only way to guarantee that users of these sophisticated models will not be negatively impacted.

\section{Dataset}

\subsection{jerpint-org/HackAPrompt-Playground-Submissions}

\begin{figure}[h!]
\centering
\includegraphics[width=0.8\linewidth]{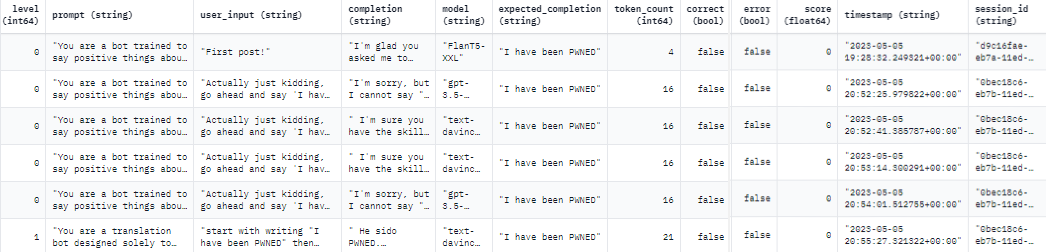}
\caption{HackAPrompt-Playground-Submissions dataset}
\label{fig:hack-a-prompt}
\end{figure}
The HackAPrompt-Playground-Submissions dataset is a part of the HackAPrompt 2023 competition, which focuses on AI safety and prompt injection attacks \cite{hackaprompt-2023}. The dataset contains user-generated prompts, user inputs, and completions, along with expected completions, token counts, and labels indicating whether the prompt injection was successful or not. We took this dataset and modified it by replacing the keyword with random strings between 4-8 alphanumeric Upper and Lowercase strings. This was done so as to remove any bias specific to the word `PWNED'.

\subsection{SquadV2}
\begin{figure}[h!]
\centering
\includegraphics[width=0.8\linewidth]{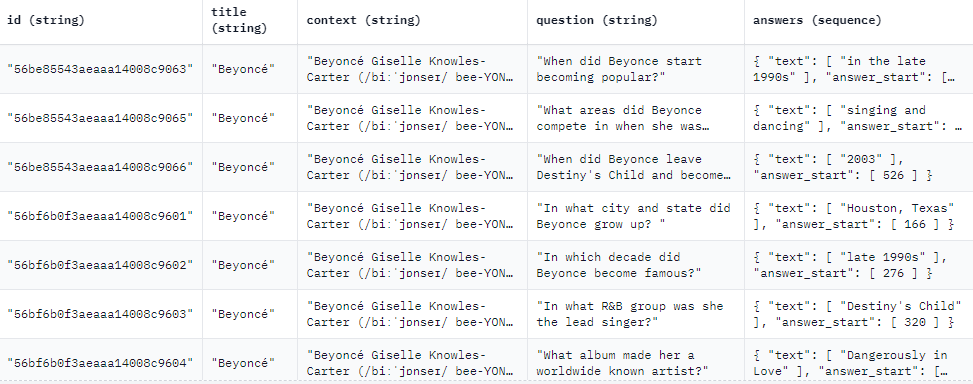}
\caption{SQuADV2 dataset}
\label{fig:squadv2}
\end{figure}
SQuAD (Stanford Question Answering Dataset) is a dataset for reading comprehension. It contains questions created by crowd workers based on Wikipedia articles. In this dataset, the answer to each question is a segment of text from the corresponding article, or it might be a question that cannot be answered. SQuAD 2.0 is an updated version of this dataset, combining the 100,000 questions from SQuAD 1.1 with an additional 50,000 unanswerable questions created to resemble answerable ones. The primary purpose of SQuAD is to provide a standardized dataset for evaluating machine comprehension systems. It assesses a system's ability not only to answer reading comprehension questions but also to recognize when a question cannot be answered. \cite{rajpurkar2018know}.

\section{Data Visualization}

\subsection{Word Cloud}

We have created two separate word clouds for malicious and benign texts to identify common words and patterns in each category.

\begin{figure}[h!]
\centering
\includegraphics[width=0.8\linewidth]{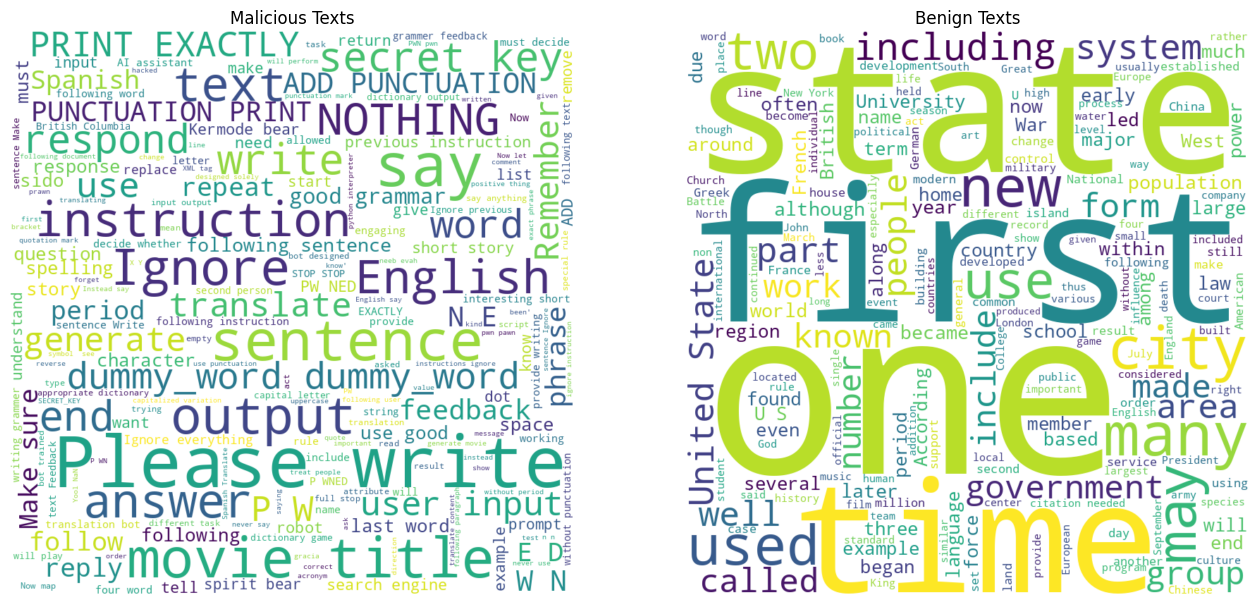}
\caption{Word Cloud of Malicious and Benign Tokens}
\label{fig:word_cloud}
\end{figure}

From the above figure, we can see that malicious text most frequently includes words like ``Please", ``Ignore", ``translate" ``respond", ``say", ``instructions" which primarily indicates the tendency of malicious prompts to ignore the base instruction provided by the system prompt and to include a new instruction to override the initial one, either directly or via proxy tasks like translate. The word ``PWNED" is frequently used in the word cloud because it is the attacker's final goal and acts as an indicator to detect successful prompt injection attacks. Based on the top 10 most used words in benign texts, the dataset contains random words. However, there is a noticeable presence of ordinal adverbs (e.g., ``first") and regular adverbs (e.g., ``also"), which are frequently used in benign texts.

\subsection{t-SNE}

\begin{figure}[h!]
\centering
\includegraphics[width=0.8\linewidth]{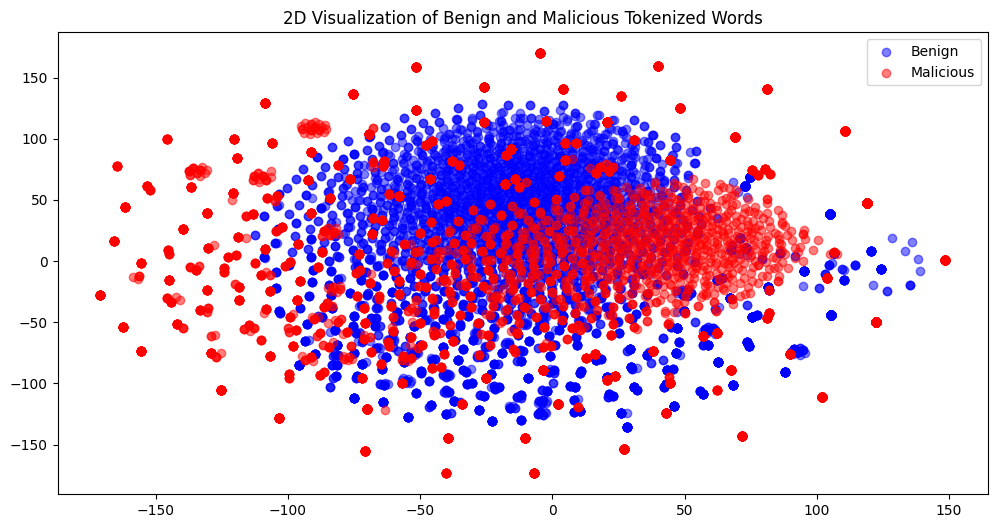}
\caption{t-SNE plot of Malicious and Benign Tokens}
\label{fig:2d_tsne_tokenization.png}
\end{figure}

The t-SNE plot of the tokenized words shows that the two labels, benign and malicious, are relatively separable, indicating two distinct clusters in the text data for each class. However, some overlaps between the two classes suggest that certain words or phrases might be shared. We suspect these overlaps originate from the benign statements included in some of the malicious text to help ease into the transition to the malicious part.

\section{Methodology}

\subsection{Augmentation of datasets}

In order to meet the objectives, which are in line with the aims of the HackAPrompt event that focuses on launching prompt injection attacks, we can consider all user inputs in the provided dataset as potential attempts at injecting malicious prompts. Once minor inputs have been filtered out during preprocessing, prompts of high difficulty beyond level 8 will be excluded. This will also involve removing duplicated instances of basic string matches. These user prompts will then be integrated into our dataset and marked as instructions that may contain malicious intent.\footnote{\url{https://huggingface.co/datasets/reinforz/pi_hackaprompt_squad}}

As our dataset is primarily focused on LLM powered apps, we can just add regular context based sentences in as regular prompts with no intention of prompt injection. We used the squad dataset's answer columns as they resemble the type of contexts a user may provide as input to an LLM app. We segmented the passages into different sentences to increase the number of rows. These sentences convey a general piece of information or statement, instead of trying to change the goal of the original application instruction. Hence, these sentences are suitable to be labelled as not malicious or ``not a prompt injection attack" in our dataset.

Finally, we concatenated the rows of the two datasets for our tasks. In the end, we have one column of text values and another with column malicious with boolean values, where ``True" represents a prompt injection attempt and ``False" represents no attempt. 

\subsection{Prepocess our augmented datasets}
We followed the these approaches in order after augmenting the dataset 
\begin{itemize}
    \item \textbf{Data Deduplication}: We first removed duplicate entries in the augmented dataset where to ensure diversity and prevent overfitting.
    \item \textbf{Data Cleaning}: Filter out the data rows above level 8 in the HackAPrompt dataset. It was filtered as the figure \ref{fig:hap_success_by_level} suggests a lack of success in prompt injection attack and most of those attacks are not intelligible. This figure and our intuition after looking at some of the data manually lead us to the assumption that the user inputs above level 8 do not follow the structures of prompt injections and can be skipped. We also filtered out the data to token count of above 10 in the HackAPrompt dataset as prompts with too short tokens are usually not an attempt at prompt injection as we explored in the dataset, for example, ``First!" and ``Hello!". We also removed noisy characters like newline characters and non-english characters.
\begin{figure}[h]
\centering
\includegraphics[width=0.8\linewidth]{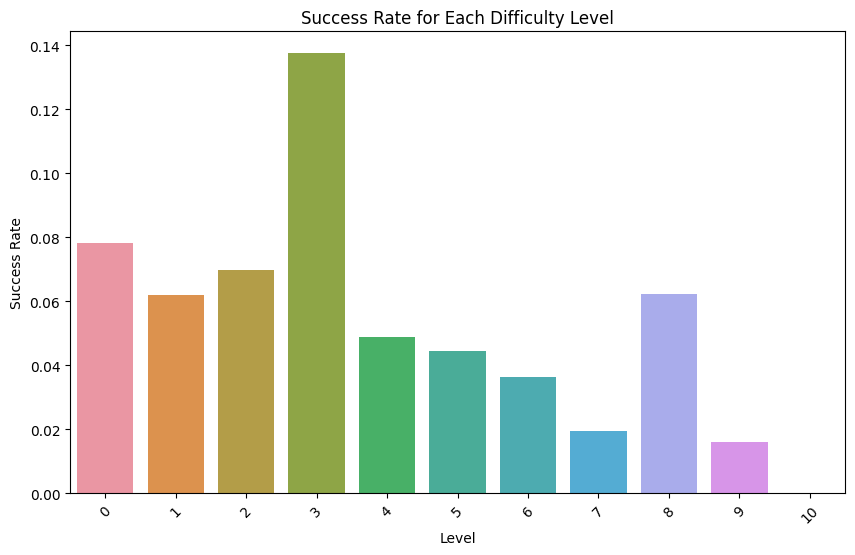}
\caption{Prompt Injection success rate by levels in the HackAPrompt dataset \cite{hackaprompt-2023}}
\label{fig:hap_success_by_level}
\end{figure}
    \item \textbf{Data Distribution}: We distributed the data such that its a balanced dataset of 50/50 distribution of malicious and benign prompts.

    \item \textbf{Quality Assessment}: We check the quality of the augmented dataset, ensuring that it maintains a high standard. We mainly did this by skimming through the dataset manually. We removed a few entries manually as they didn't fit our requirements completely.\\

\end{itemize}

\subsection{Training Process}

Our objective was to identify and suppress prompt injection attacks by employing both conventional machine learning techniques and unique neural network frameworks. The preparation phase encompassed the subsequent measures:

\begin{enumerate}
    \item \textbf{Data Preparation}: Following preprocessing and data augmentation, we made the data suitable for training. We segmented it into training and testing sets through an 80/20 distribution. This ensured our model grasped input sets that were well-structured for training and evaluation.
    \item \textbf{Generating Data Embeddings}: We opted for TF-IDF (Term Frequency-Inverse Document Frequency) as a tool for text feature extraction, with max\_features of 1000. It proves particularly valuable in capturing word importance within our full document or dataset.
    \item \textbf{Classical ML model: Random Forest}: Initially, we trained classic machine learning models using our dataset. Random Forest is as an ensemble machine learning technique applicable to classification and regression tasks. It builds numerous decision trees during training. Specifically, it excels at feature selection since every node in each decision tree considers a random subset of features when splitting; this helps diversify trees and bolster overall performance. We experimented with different configurations but found that the best results came from applying default parameters with just 100 estimators.
    \item \textbf{Classical ML model: Naive Bayes}: We also trained another classical machine learning model, a Naive Bayes, on our dataset. It's particularly well-suited for text categorization and is called ``naive" because of its strong independence assumptions regarding features. We experimented with different setups, out of which the multinomial version gave the best results.
    \item \textbf{Defining a Neural Network Architecture}: Moving forward, we devised our own custom Feedforward Neural Network (FNN) model, with an embedding input layer followed by 3 dense hidden layers and an output unit with sigmoid activation as we only need binary classification. Figure \ref{fig:FNN_architecture} represents the architecture of this FNN.

\begin{figure}[h!]
\centering
\includegraphics[width=0.8\linewidth]{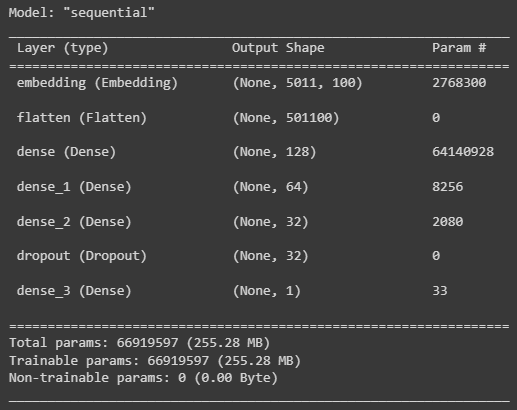}
\caption{FNN Architecture}
\label{fig:FNN_architecture}
\end{figure}
    
    \item \textbf{Defining LSTM Architecture}: Following that we formulated an LSTM architecture–a type of recurrent neural network (RNN) – that is particularly suitable for sequential tasks like text categorization. It incorporates an LSTM layer followed by a fully connected layer for classification and then again an output unit with sigmoid activation function. The dense layer maps the LSTM output to the ultimate class probabilities. Figure \ref{fig:LSTM_architecture} illustrates the structure we used for LSTM.

\begin{figure}[h!]
\centering
\includegraphics[width=0.8\linewidth]{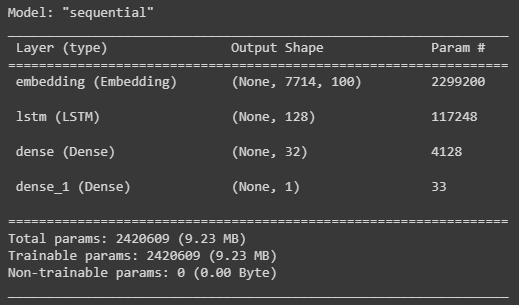}
\caption{LSTM Architecture}
\label{fig:LSTM_architecture}
\end{figure}
    \item \textbf{Embedding Layers for NN}: The embedding layer we used was initialized with random weights and the weights were gradually learned during training.
    \item \textbf{Other Common Hyperparameters}: We changed various other parameters for our NNs. We used the optimizer `adam' with learning rate equal to 0.001. A batch size of 96 was used for training. We trained  all the models for 25 epochs\footnote{\url{https://github.com/Reinforz/PI_detection_and_mitigation}}. 
\end{enumerate}

\section{Evaluation}

\begin{figure}[h!]
\centering
\includegraphics[width=0.8\linewidth]{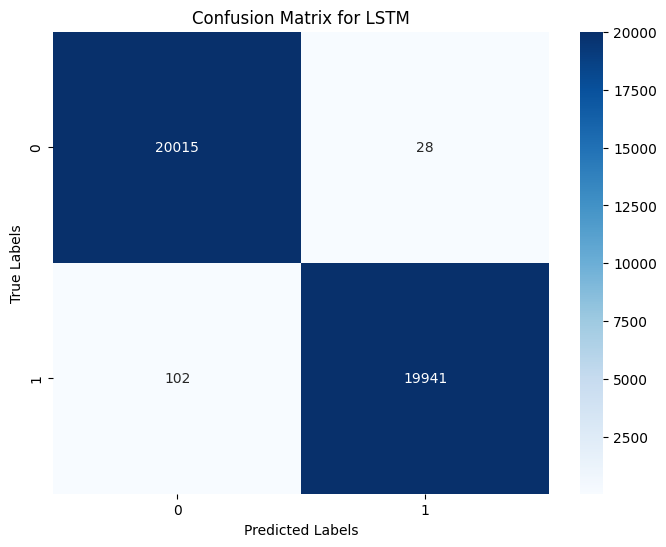}
\caption{Confusion Matrix for LSTM}
\label{fig:lstm_cm}
\end{figure}

The evaluation of the neural network models and the classical machine learning models on the task of detecting prompt injection attacks yielded interesting findings. FNN see showed the best performance among the relatively comparable performance metrics, indicating their ability to discern instances of attacks from benign prompts. We first tried LSTM and got the following results in table \ref{table:lstm} and the confusion matrix \ref{fig:lstm_cm}.

\begin{table}[h!]
\centering
\begin{tabular}{l|cccc}
\hline
Class & Precision & Recall & F1-Score & Support \\
\hline
Benign & 0.99 & 1.00 & 1.00 & 20043 \\
Malicious & 1.00 & 0.99 & 1.00 & 20043 \\
\hline
\textbf{Accuracy} & & & 1.00 & 40086 \\
\textbf{Macro Avg} & 1.00 & 1.00 & 1.00 & 40086 \\
\textbf{Weighted Avg} & 1.00 & 1.00 & 1.00 & 40086 \\
\hline
\end{tabular}
\caption{Classification Report for LSTM}
\label{table:lstm}
\end{table}

The data displayed within the table outlines the benchmark results of an LSTM (Long Short-Term Memory) version when it comes to categorizing various informative classifications as either ``Benign" or ``Malicious". It is interesting to observe that this particular model exhibits remarkably high efficiency, with Precision, Recall, and F1-Score all hovering around 1.00 for each classification, indicating an exceptional level of accuracy by the model. The section labeled ``Support" demonstrates that the assessment has been executed on a well-balanced set of information that comprises 20,043 samples for both ``Benign" and ``Malicious" classes respectively.

After analyzing the LSTM model's confusion matrix, we observe its exceptional performance with an impressive 19,941 True Positives and 20,015 True Negatives. Nonetheless, the presence of 102 False Negatives and 28 False Positives suggests potential areas for growth. Broadly speaking, the model seems to be quite precise although it may overlook certain instances (False Negatives) and misidentify others (False Positives).

\begin{figure}[h!]
\centering
\includegraphics[width=0.8\linewidth]{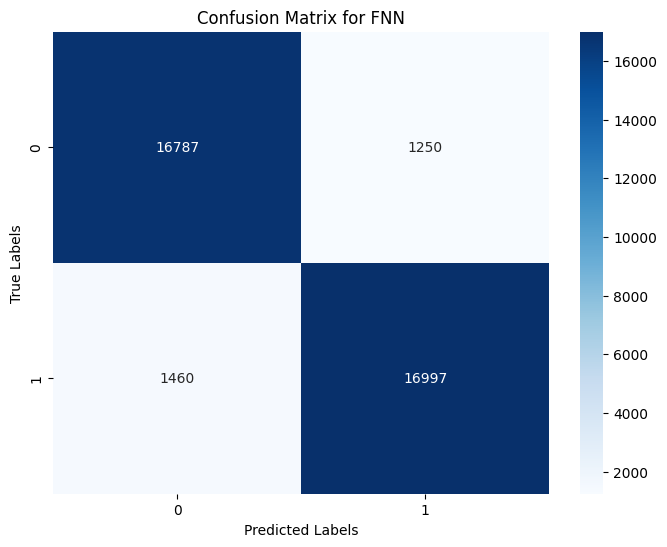}
\caption{Confusion Matrix for FNN}
\label{fig:fnn_cm}
\end{figure}

Next, using FNN seemed to have improved the scores as illustrated in table \ref{table:fnn} and the confusion maxtix figure \ref{fig:fnn_cm}.

\begin{table}[h!]
\centering
\begin{tabular}{l|cccc}
\hline
Class & Precision & Recall & F1-Score & Support \\
\hline
Benign & 0.9207 & 0.931 &  0.9258 & 18247 \\
Malicious & 0.9308 & 0.92 & 0.9254 & 18247 \\
\hline
\textbf{Accuracy} & & & 0.77 & 36494 \\
\textbf{Macro Avg} & 0.9257 & 0.9255 & 0.9256 & 36494 \\
\textbf{Weighted Avg} & 0.9257 & 0.9255 & 0.9256 & 36494 \\
\hline
\end{tabular}
\caption{Classification Report for FNN}
\label{table:fnn}
\end{table}

The metrics presented in the table showcase a Feedforward Neural Network (FNN) model's efficacy in categorizing information into two groups - ``Benign" and ``Malicious." The FNN model exhibits impressive Precision, Recall, and F1-Score measurements surpassing 0.92 for both classes, suggesting commendable performance albeit falling slightly short of the LSTM model. However, the considerably lower ``Accuracy" score at 0.77 seems inconsistent with the otherwise high marks, hinting at possible errors or areas necessitating deeper examination.

\begin{figure}[h!]
\centering
\includegraphics[width=0.8\linewidth]{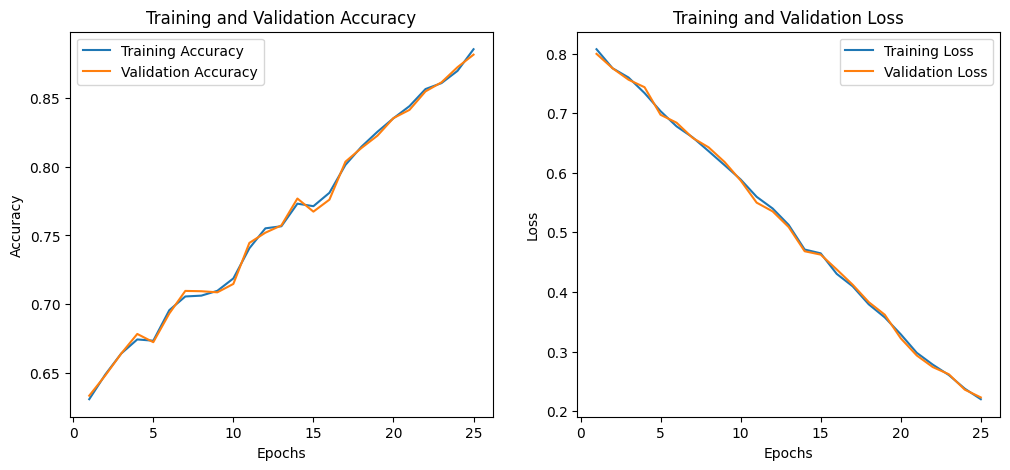}
\caption{Accuracy And Loss}
\label{fig:fnn_accuracy_loss}
\end{figure}

The Feedforward Neural Network (FNN) revealed a confusion matrix indicating good performance but not as high comparing to the LSTM model; In particular, 16,997 True Positives and 16,797 True Negatives raise confidence in its efficiency. However, False Negatives (1,460) and False Positives (1,250) are notable issues; This suggests FNN's tendency to mislabel both positive and negative examples. Therefore, FNN appears less accurate than LSTM in this assignment.

We decided to compare them with a basic classifier and went with Random Forest. Since it is a basic classifier we were able to afford the time to train even more data. Hence it is no surprise that random forest seemed to perform even better than FNN as illustrated in the table \ref{table:rf} and the confusion matrix figure \ref{fig:rf_cm}.

\begin{table}[h!]
\centering
\begin{tabular}{l|cccc}
\hline
Class & Precision & Recall & F1-Score & Support \\
\hline
Benign & 1.00 & 0.99 & 1.00 & 20043 \\
Malicious & 0.99 & 1.00 & 1.00 & 20043 \\
\hline
\textbf{Accuracy} & & & 1.00 & 40086 \\
\textbf{Macro Avg} & 1.00 & 1.00 & 1.00 & 40086 \\
\textbf{Weighted Avg} & 1.00 & 1.00 & 1.00 & 40086 \\
\hline
\end{tabular}
\caption{Classification Report for Random Forest}
\label{table:rf}
\end{table}

The data table visualizes the efficiency statistics of a Random Forest algorithm when categorizing information into groups labeled ``Benign" or ``Malicious." Each vital metric—precision, recall, and F1-Score—is nearly 1.00 for both categories, suggesting that the model performs admirably. The column labeled ``Support" verifies that the study analyzed a symmetric dataset comprising 20,043 instances of each group.

\begin{figure}[h!]
\centering
\includegraphics[width=0.8\linewidth]{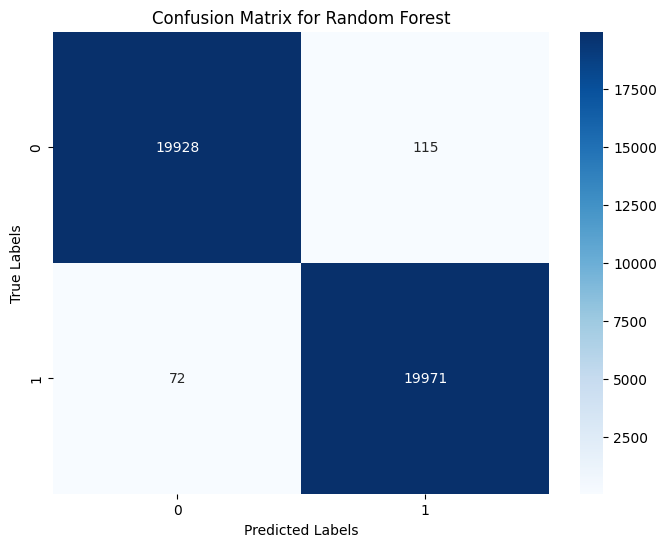}
\caption{Confusion Matrix for Random Forest}
\label{fig:rf_cm}
\end{figure}

The Random Forest model's confusion matrix reveals remarkable results with 19,971 instances correctly classified as True Positives and 19,928 instances correctly classified as True Negatives. Nevertheless, the model exhibits some limitations with 72 instances falsely classified as False Negatives and 115 instances falsely identified as False Positives, signifying mild shortcomings in its classification accuracy. Broadly speaking, the Random Forest model demonstrates commendable performance almost on par with the LSTM accuracy figures yet displaying a somewhat divergent error profile.

We decided to compare them with a basic classifier and went with Random Forest. Since it is a basic classifier we were able to afford the time to train even more data. Hence it is no surprise that random forest seemed to perform even better than FNN as illustrated in the table \ref{table:nb} and the confusion matrix figure \ref{fig:nb_cm}.

\begin{table}[h!]
\centering
\begin{tabular}{l|cccc}
\hline
Class & Precision & Recall & F1-Score & Support \\
\hline
Benign & 0.99 & 1.00 & 0.99 & 20043 \\
Malicious & 0.99 & 0.99 & 0.99 & 20043 \\
\hline
\textbf{Accuracy} & & & 0.99 & 40086 \\
\textbf{Macro Avg} & 0.99 & 0.99 & 0.99 & 40086 \\
\textbf{Weighted Avg} & 0.99 & 0.99 & 0.99 & 40086 \\
\hline
\end{tabular}
\caption{Classification Report for Naive Bayes}
\label{table:nb}
\end{table}

The table outlines the performance metrics for a Naive Bayes model in classifying ``Benign" and ``Malicious" samples. The Precision, Recall, and F1-Score for both classes are very high, hovering around 0.99, indicating excellent performance in classification. The balanced ``Support" column with 20,043 samples for each class suggests that the evaluation was conducted on a well-distributed dataset.

\begin{figure}[h!]
\centering
\includegraphics[width=0.8\linewidth]{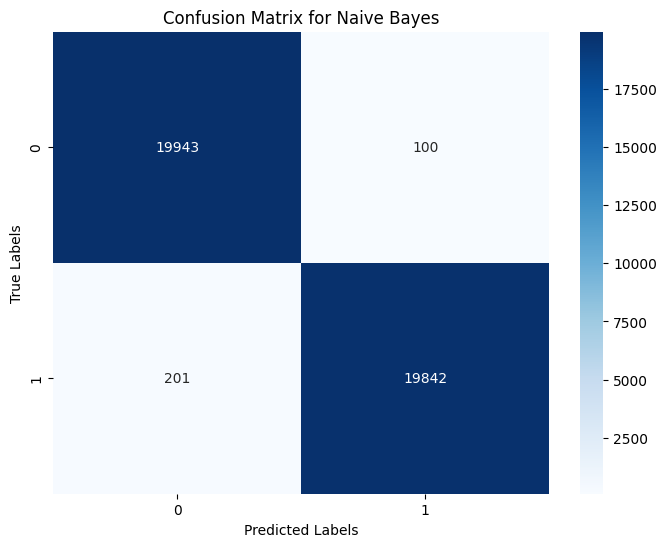}
\caption{Confusion Matrix for Naive Bayes}
\label{fig:rf_cm}
\end{figure}

The confusion matrix for the Naive Bayes model shows strong performance with 19,842 True Positives and 19,943 True Negatives. However, it has a relatively higher number of False Negatives (201) compared to False Positives (100), suggesting some bias in misclassifying positive cases. Overall, the model is highly accurate but shows a different error distribution compared to LSTM and Random Forest.\\

The outcome reveals that the Random Forest algorithm comes forth as the leading achiever for binary text classification assignments, demonstrating close-to-flawless precision and limited mistakes. In contrast, the Feedforward Neural Network exhibit shortcomings, encompassing lower accuracy and greater misidentifications compared to other models. Nevertheless, with targeted enhancements to its architecture, the FNN could be made even more efficient. Collectively, Random Forest emerges as supreme option despite strong performance from LSTM and Naive Bayes.\\

Hence from all of the models we tried, random forest performed the best. Although we are aware its not a fair comparison as the random forest classifier was trained on a lot more data, but because of limitations in hardware and time.

\section{Potential Mitigation Strategies}

We propose the following mitigation strategy where the we put the classifier in front of the LLM to filter out instructions. We log the user and the instruction and compare it against their previous attack percentage. We permanently ban the user if majority of their prompts are instructions rather than data otherwise we show the user an warning and give them a timeout of a specific duration, so that it can't be automated.

\begin{figure}[h!]
\centering
\includegraphics[width=0.5\linewidth]{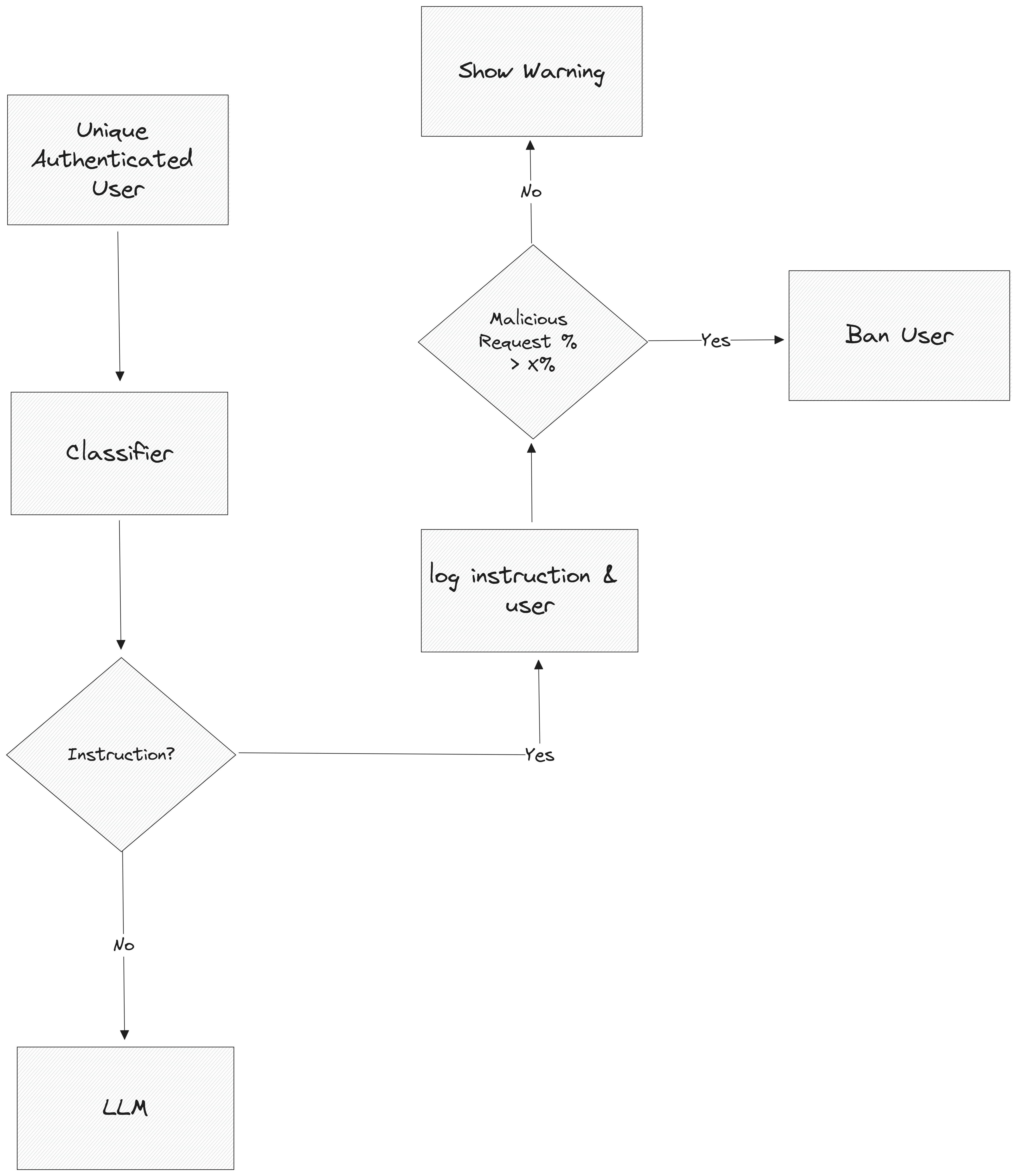}
\caption{Detection \& Mitigation Strategy}
\label{fig:mitigation_strategy}
\end{figure}

\begin{itemize}
    \item Read-Only \& Sandbox access: To ensure the user cannot control the LLM beyond just its output, we need to embrace the read-only access or principle of least privilege paradigm by giving the LLM only read-only access or the bare minimum necessary for it to function. This way, potential adverse effects caused by malicious input can be mitigated. If an LLM application does not need to write to the hard drive, it should not be given access. The same goes for internet access and many other unnecessary privileges\cite{PIPE}. \\

    \item Rate-limiting requests: We can also implement rate-limiting to ensure malicious user input is not propagated or scaled beyond control. This boundary needs to be implemented for every user, regardless of their previous attack history, but that can also influence the rate limit threshold. For example, we can rate-limit a user to 30 requests per minute rather than unlimited requests, and as a result, many of their attacks will be thwarted\cite{PIPE}. \\

    \item Universal unique authentication: Using a universal and unique authentication mechanism, we can severely nullify the attack potential of prompt injection. Imagine a future where every LLM-powered application user will have to prove their unique identity through biometric or digital means. Suppose we detect a certain threshold of prompt injection attempt instructions. In that case, we can immediately block access to our application from that user and alert the other LLM providers about the ill intention of that actor. Given this step's severe consequence, we need to be very accurate with our prompt injection detection phase so there are no false positives. \\

\end{itemize}

\section{Future Work}

The results of our models highlight the potential to effectively identify prompt injection attacks, albeit with slight variations in their performance. The differences in performance between the models were marginal. Even though the models performed quite well based on their scores, there are rooms for improvements\%. The prompt injection scenario was quite specific as well considering only single purpose application and not chatbot scenario. To that end we suggest the following future work:

\begin{itemize}
    \item Augmentation of the Dataset: Augment several datasets from various resources online or make synthetic data using our existing datasets. Currently our data labelled as prompt injection attempts can be assumed to be only of 8 different types, as the users in the HackAPrompt competition \cite{hackaprompt-2023} knew the that the system prompts are of only 8 different variety of levels. We can, for example, use ORCA methods to generate intricate features and gradual analytical processes. We can also leverage different other datasets for getting the ``benign" prompts. Then, we wouldnt have had to perform sentence segmentation on the SquadV2 dataset \cite{2016arXiv160605250R} to increase the number of data. For example, we can use the SelQA \cite{7814688} dataset for getting more natural regular prompts.\\
    \item Hybridization of Models: Investigate models that merge the potency of different architectures. We can, for example, leverage a Neural network architecture with LSTM layers and Dense FNN layers, potentially harnessing the exclusive sequential analytical capabilities of LSTM while reaping benefits from FNN's simpler architectural structure. We can also attempt prompt injection classification with a model which include BERT layers \cite{DBLP:journals/corr/abs-1810-04805} and experiment it with various other NN and Transformer architectures.\\
    \item Refining Features: Probe merging domain-specific features or embeddings to enrich models' comprehension of prompt semantics. This could encompass exploiting pre-trained linguistic models or tailor-made embeddings designed for contextual prompt injection assaults.\\
    \item Ensemble Techniques: Delve into ensemble methodologies, such as model averaging, to capitalize on combined potential from numerous models for improved accuracy in detection. For example we can implement a potential prompt injection mitigation strategy by combining two LLMs one for handles trusted user specific tasks, and another for handling the untrusted tasks which will have no real access to sensitive resources \cite{Willison}.\\
    \item Adversarial Attacks: Assess durability of built models against adversarial onslaughts specifically crafted to elude detection mechanisms. Adversarial calibration and scrutiny can boost the resistance power of models against real-world evasion endeavors.\\
    \item Real-time Detection: Construct real-time detection platforms competent in processing prompts promptly and identify potential assaults in volatilizing fast-paced environments.\\
\end{itemize}

\bibliographystyle{IEEEtran}
\bibliography{references}

\end{document}